\documentclass[twocolumn,amsmath,amssymb,floatfix, superscriptaddress,prb]{revtex4-2}
\usepackage{lineno,hyperref}
\usepackage{bm}
\usepackage{amsmath}
\usepackage{gensymb}
\usepackage{epstopdf}
\usepackage{booktabs}
\usepackage{longtable}
\usepackage{tabularx}
\usepackage{multirow}
\usepackage{setspace}
\usepackage{mathrsfs}

\usepackage{graphicx}
\usepackage{ulem}
\usepackage{xcolor}
\usepackage{color}
\usepackage{hyperref}

\begin{document}

\title{Large saturation moment and high ferromagnetic transition temperature in a structurally disordered inverse Heusler alloy Fe$_2$RuGe}

\author{Sudip Chakraborty}
\affiliation{Condensed Matter Physics Division, Saha Institute of Nuclear Physics,
              1/AF Bidhannagar, Kolkata 700064, India}
\affiliation{Homi Bhabha National Institute, Anushaktinagar, Mumbai 400094, India}
\author{Shuvankar Gupta}
\affiliation{Condensed Matter Physics Division, Saha Institute of Nuclear Physics,
               1/AF Bidhannagar, Kolkata 700064, India}
\affiliation{Homi Bhabha National Institute, Anushaktinagar, Mumbai 400094, India}
\author{Vidha Bhasin}
\affiliation{Atomic \& Molecular Physics Division, Bhabha Atomic Research Centre, Mumbai 400 094, Maharashtra,  India}
\affiliation{Homi Bhabha National Institute, Anushaktinagar, Mumbai 400094, India}
\author{Santanu Pakhira}
\affiliation{Ames National Laboratory, Iowa State University, Ames, Iowa 50011, USA}
\author{C\'{e}line Barreteau}
\affiliation{Univ. Paris Est Creteil, CNRS, ICMPE, UMR 7182, 2 rue Henri Dunant, 94320 Thiais, France}
\author{Jean-Claude Crivello}
\affiliation{Univ. Paris Est Creteil, CNRS, ICMPE, UMR 7182, 2 rue Henri Dunant, 94320 Thiais, France}
\author{Shambhu Nath Jha}
\affiliation{Atomic \& Molecular Physics Division, Bhabha Atomic Research Centre, Mumbai 400 094, Maharashtra,  India}
\author{Dibyendu Bhattacharyya}
\affiliation{Atomic \& Molecular Physics Division, Bhabha Atomic Research Centre, Mumbai 400 094, Maharashtra,  India}
\author{Maxim Avdeev}
\affiliation{Australian Nuclear Science and Technology Organisation, Locked Bag 2001, Kirrawee DC, NSW 2232, Australia}
\affiliation{School of Chemistry, The University of Sydney, Sydney, NSW 2006, Australia}
\author{Val\'{e}rie Paul-Boncour}
\affiliation{Univ. Paris Est Creteil, CNRS, ICMPE, UMR 7182, 2 rue Henri Dunant, 94320 Thiais, France}
\author{Jean-Marc Greneche}
\affiliation{Institut des Mol\'{e}cules et Mat\'{e}riaux du Mans, IMMM UMR CNRS 6283, Le Mans Cedex 9, 72085, France}
\author{Eric Alleno}
\affiliation{Univ. Paris Est Creteil, CNRS, ICMPE, UMR 7182, 2 rue Henri Dunant, 94320 Thiais, France}
\author{Chandan Mazumdar}
\email{chandan.mazumdar@saha.ac.in}
\affiliation{Condensed Matter Physics Division, Saha Institute of Nuclear Physics,
               1/AF Bidhannagar, Kolkata 700064, India}
\affiliation{Homi Bhabha National Institute, Anushaktinagar, Mumbai 400094, India}

\date{\today}

\begin{abstract}
We report the successful synthesis of a new 4$d$ based polycrystalline inverse Heusler alloy Fe$_2$RuGe by an arc melting process and have studied in detail its structural, magnetic and transport properties complemented with first principle calculations.
X-ray and neutron diffraction, Extended X-ray Absorption Fine Structure and $^{57}$Fe M\"{o}ssbauer spectroscopic studies confirm the single phase nature of the system where the Fe and Ru atoms are randomly distributed in the 4$c$ and 4$d$ Wyckoff positions in a ratio close to 50:50. The formation of the disordered structure is also confirmed by the theoretical energy minimization calculation. Despite the random cross-site disorder of Fe and Ru atoms, magnetic measurements suggest not only a high Curie temperature of $\sim$860\,K, but also a large saturation magnetic moment $\sim$4.9\,$\mu_B$ per formula unit at 5\,K, considerably exceeding the theoretical limit (4\,$\mu_B$ per formula unit) predicted by the Slater-Pauling rule.
Only a few Fe-based inverse Heusler alloys are known to exhibit such high Curie temperatures. Neutron diffraction analysis coupled with the isothermal magnetization value indicates that the magnetic moments in Fe$_2$RuGe are associated with Fe-atoms only, which is also confirmed by Mössbauer spectrometry. Interestingly, in comparison to the cubic or hexagonal phase of the parent compound, Fe$_3$Ge, the Curie temperature of Fe$_2$RuGe has increased significantly despite the substitution of the nonmagnetic, yet isoelectronic element Ru in this structurally disordered compound. Our theoretical calculation reveals that the large Fe moment ($\sim2.8\mu_B$/Fe) on the 4$b$ site can be attributed to a charge transfer from this Fe site towards its Ru neighbours. Such a substantial increase in magnetic moment due to electron charge transfer has not previously been reported in a Heusler alloy system.
\end{abstract}

\maketitle

\section{Introduction}

Heusler alloys (HA) are known to be a series of enigmatic compounds that are often known to exhibit high ferromagnetic transition temperatures, even when all individual constituents are nonmagnetic~\cite{graf2011simple,felser2015heusler}.
HAs have a general chemical formula of $X_2YZ$, where $X$ and $Y$ are transition metal atoms and $Z$ is generally a main group element atom.
They crystallize in the Cu$_2$MnAl-type or $L2_1$-type structure ($Fm\bar{3}m$).
Most HAs are highly ordered cubic intermetallic compounds with three  crystallographic positions in their crystal lattice, viz. 4$a$ (0,0,0), 4$b$ ($\frac12$,$\frac12$,$\frac12$) and 8$c$ ($\frac14$,$\frac14$,$\frac14$)~\cite{graf2011simple}.
It is generally seen that the occupancy of an atom at a particular site is determined by its number of valence electrons.
Those with more valence electrons ($X$) tend to occupy the 4$c$ and 4$d$ positions, while atoms with fewer valence electrons ($Y$) tend to occupy the 4$b$ position, which is generally known as the valence electron number rule~\cite{mondal2018ferromagnetically}.
However, there is also a subclass of HAs, where $Y$ and one of the $X$ atoms interchange their occupancy, and such systems are known as inverse HA, forming in the XA-type structure, also known as the Hg$_2$CuTi structure~\cite{gasi2013iron,du2013crossover}.
This latter crystallises in $F\bar{4}3m$ space group with a splitting of $8c$ sites from $Fm\bar{3}m$ into two individual $4c$ and $4d$ sites.
Another fascinating feature of HA system is the structural tunability, as many different combinations of 3$d$, 4$d$ and 5$d$ transition metal elements can be used in the $X$ and $Y$ positions. Interestingly, one may realize that the choice of constituent elements and their occupancies determines of the electronic band structure, which in turn controls most of the physical properties of these materials. As a result, HAs remain a very attractive system for appropriate crystal engineering~\cite{fu2015band,chauhan2020defect,chauhan2019compositional,ozdougan2008engineering}.

The high Curie temperature and large saturation magnetic moment displayed by Heusler alloys recently got renewed attention due to their multi-functional properties viz. Half metallic ferromagnetism (HMF)~\cite{de1983new,wurmehl2005geometric,gupta2022coexisting}, semi-metallic behavior~\cite{mondal2018ferromagnetically}, magnetic shape memories~\cite{siewert2012first,chopra2000magnetic}, magnetic spin valve effects~\cite{huang2013simulation}, topological insulators~\cite{lin2010half,chadov2010tunable}, magnetic skyrmions~\cite{zuo2018zero,giri2020robust}, spin-gapless semiconductors~\cite{ouardi2013realization,bainsla2015spin}, etc., as all these properties could be used in many practical applications developed recently. For example, spintronics devices are considered to have huge advantages over conventional electronic devices due to their high speed for data processing, non-volatile nature and reduced power consumption rate~\cite{vzutic2004spintronics}. It is generally seen that HA with the combination of both Co and Fe is particularly good for obtaining a high magnetic moment and elevated Curie temperature. A substantial number of Co-based systems are particularly known to exhibit HMF behaviour and could be used for spintronic-related applications~\cite{wurmehl2005geometric}~\cite{telling2006interfacial}. On the other hand, although Fe-based HAs are rarely known to show such features, they have an enormous potential to be used as economically viable rare-earth free permanent magnet.
In the quest for such new permanent magnetic material, the iron-rich compound, Fe$_3$Ge turns out to be one of the most remarkable systems~\cite{mcguire2018tuning}. The compound could be considered as the binary equivalent of Heusler alloy, where both the $X$ and $Y$ atoms are identical.
Very interestingly, the compound is found to be stable at room temperature in two polymorphic crystal structures, viz., a cubic $L2_1$ phase (Cu$_3$Au type, $Pm\bar{3}m$, space group no. 221) formed by annealing at low temperatures ($<$ 1073\,K) and a hexagonal $D0_{19}$ phase (Mg$_3$Cd type, $P6_3/mmc$, space group no. 194) stabilized when the system is annealed at higher temperatures ($>$1073\,K)~\cite{mcguire2018tuning}.
Although both phases exhibit high Curie temperatures (655\,K for the hexagonal phase and 755\,K for the cubic phase) and large magnetic moments ($\sim6\mu_B$/f.u.) at room temperature, the anisotropic magnetic properties of the hexagonal phase have generated considerable interest in its use as a permanent magnet.
A lot of attempts have been made to modify the magnetic anisotropy by partially replacing Fe atoms with other 3$d$ elements, viz., V, Cr, Mn, Co and Ni~\cite{mahat2020tuneable,mahat2021influence,keshavarz2019fe2mnge,gasi2013iron}. It has been found that depending on the use of heavier or lighter 3$d$
element substitution, the structural as well as magnetic properties can vary widely. Both Fe$_2$CoGe and Fe$_2$NiGe form in cubic inverse Heusler alloy structure~\cite{gasi2013iron}, while  Fe$_{3-x}$Cr$_x$Ge ($x <$ 0.70) and Fe$_2$MnGe retain the hexagonal crystal structure~\cite{mahat2021influence}~\cite{keshavarz2019fe2mnge}.
Fe$_{3-x}$V$_x$Ge remains hexagonal for $x <$ 0.125, but changes to cubic for 0.125 $< x <$ 0.75~\cite{mahat2020tuneable}. The ferromagnetic Curie temperature for Fe$_2$CoGe ($T_{\rm C}$ = 925\,K) and Fe$_2$NiGe ($T_{\rm C}$ = 760\,K) is also higher than that of both the hexagonal as well as cubic phases of the parent compound, Fe$_3$Ge.
In addition, although Co and Ni are considered magnetic elements, the magnetic moment is confined only to Fe in Fe$_2$NiGe, while both Fe and Co carry magnetic moment in Fe$_2$CoGe. On the other hand, the Curie temperature decreases in case of substitution by 3$d$ elements lighter than Fe~\cite{mahat2020tuneable,mahat2021influence,keshavarz2019fe2mnge,gasi2013iron}.

In light of such contrasting magnetic properties of Fe$_{3-x}T_x$Ge depending on the electronic configuration of 3$d$ $T$-element, it would be interesting to understand the role of the valence electron and the  electronic band structure in this system. The substitution of isoelectronic elements belonging to the 4$d$ (Ru) or 5$d$ (Os) series would be highly helpful, despite the fact that both are non-magnetic and isoelectronic to Fe, and are also susceptible to structural  disorder.
In this work, we report the successful synthesis of a new inverse Heusler alloy Fe$_2$RuGe and explored its structural and physical properties through different experimental and theoretical techniques.
Two different spectroscopic techniques, viz., $^{57}$Fe M\"{o}ssbauer spectrometry and Extended X-ray Absorption Fine Structure (EXAFS) have been adopted to investigate the local structural disorders, While neutron diffraction studies have been carried out to investigate the magnetic spin arrangements.
The electronic band structure calculation is also presented to complement the experimental results.

\section{Methods}
\subsection {Experimental details}
The Fe$_2$RuGe polycrystalline ingot was prepared by the standard arc melting method in argon atmosphere on water-cooled Cu hearth. A stoichiometric amount of Fe chunk ($>$99.9\%), Ru powder ($>$99.9\%) and Ge lump ($>$99.999\%) were used as starting materials. All the elements were melted and re-melted together for 5 times after flipping each time to improve homogeneity. A negligible amount ($<$0.6\%) of weight loss was recorded after the melting process. X-ray diffraction (XRD) measurement at room temperature was carried out using a commercial diffractometer (rotating Cu anode, 9 kW, Model: TTRAX-III, Rigaku Corp., Japan).
The single-phase nature of the as-cast sample was confirmed by Rietveld refinement using FULLPROF software~\cite{rodriguez1993recent}. Microstructural analysis and compositional homogeneity were further investigated using a wavelength dispersive electron probe microanalysis (EPMA) spectrometer [Model: SX 100, Cameca, France].

Extended X-ray Absorption Fine Structure (EXAFS) experiments on Fe$_2$RuGe were carried out at the Indus-2 Synchrotron Source (2.5 GeV, 100 mA) at the Energy-Scanning EXAFS beamline (BL-9) at Raja Ramanna Centre for Advanced Technology (RRCAT), Indore, India which operates in the energy range of 4 - 25\,keV. Both transmission and fluorescent modes can be used to perform EXAFS measurements in this beamline. A Rh/Pt coated collimating meridional cylindrical mirror is used and the collimated beam is reflected by a mirror that is monochromatized by a double crystal monochromator (DCM) of Si(111) (2\textit{d} = 6.2709\,\AA). For horizontal focusing, a sagittal cylinder (the second crystal of DCM) is  used, and for vertical focusing of the beam, a Rh/Pt coated bendable post mirror facing down at the sample position is employed. The higher harmonics content of the X-ray beam is rejected by detuning the second crystal of  DCM. For EXAFS measurements, we have used the fluorescence mode. In this method, the incident X-ray beam angle is 45 degrees to the sample position, and a fluorescence detector, positioned at 90 degrees to the incident X-ray beam, is used to collect the signal. To measure the incident flux (\textit{I}$_0$), an ionization chamber detector is placed before the sample, and  fluorescence intensity (\textit{I}$_f$) is measured by a fluorescence detector.
The relation, $\mu$ = ($I_f$/$I_0$), is used to calculate the X-ray absorption coefficient of the sample. The spectrum (as a function of energy) was acquired by scanning the monochromator over a particular range. In order to extract qualitative information about the local structure, the oscillations of the normalized absorption spectra ($\mu(E)\,vs.\,E$) were transformed to absorption function $\chi(E)$ using the relation:
\begin{equation}
\chi(E) = \frac{\mu(E)-\mu_{0}(E)} {\Delta\mu_{0}(E_{0})}
\label{ex1}
\end{equation}
\noindent
here, $E_0$ is the absorption edge energy, $\mu_0(E_0)$ is the background of exposed atom and $\Delta\mu_{0}(E_{0})$ is the step in $\mu(E)$ value at the absorption edge. The absorption coefficient as a function of energy $\chi(E)$, was changed to the absorption coefficient as a function of wave number $\chi(K)$ by the following equation,

\begin{equation}
K=\sqrt {\frac {2m(E-E_{0})}{\hbar^{2}}}
\label{eq2}
\end{equation}
\noindent
here, $m$ is the mass of the electron. To intensify the oscillation at high $\chi(K)$, it is multiplied by $k^2$ and then to obtain the $\chi(R)\,vs.\,R$ plots, a Fourier transformation of $\chi(k)k^2$ functions into $R$~space, regarding to the real distances calculated from the centre of the absorbing atom is done. The set of EXAFS data was analyzed by the program available in the  Demeter software package. Background reduction, Fourier transformation of the absorption spectra to derive $\chi(R)\,vs.\,R$ plots (using ATHENA code)~\cite{ravel2005athena}, simulation of the theoretical plots from the assumed crystallographic structure, and finally the fitting of experimental data using the theoretically generated spectra (by ARTEMIS code) were also included in this package~\cite{ravel2005athena}.

Isothermal magnetisation measurements at low temperatures were performed in commercially available SQUID VSM (Quantum Design Inc., USA). Magnetic measurements in the temperature range 300 - 900\,K were done in a high-temperature VSM equipped with an electromagnet (Model EV9, MicroSense, LLC Corp., USA).

Room temperature, as well as high-temperature neutron diffraction experiments ($\lambda$ = 2.4395\,\AA), were performed at ECHIDNA beam line in ANSTO, Sydney, Australia~\cite{avdeev2018echidna}.
The details of the crystal structure as well as the analysis of the magnetic structure were performed by Rietveld refinement of neutron diffraction data using the FULLPROF software.

$^{57}$Fe transmission M\"{o}ssbauer spectrometry was used to study the hyperfine structure at Fe sites. The spectra were obtained at 300\,K and 77\,K in a bath cryostat and using an electromagnetic transducer with a triangular velocity nature and a $^{57}$Co source in a Rh matrix. The sample consists of a thin powdered and homogeneous layer containing about 5 mg-Fe/cm$^2$.
The least-squares fitting method, which assumes quadrupolar doublets and/or magnetic sextets involving Lorentzian lines, has been applied to model the hyperfine structures using the MOSFIT program. The refined isomer shift values are corrected to that of $\alpha$-Fe at 300\,K, and an $\alpha$-Fe standard was used to calibrate the velocity.

Resistivity experiments using a standard four-probe method were performed in the temperature range 2 - 300\,K using a PPMS Evercool-II instrument, (Quantum Design Inc., USA).

\subsection {Computational methods}
\label{sec:DFT}
For the modelling part, the $L2_1$, inverse XA and disordered structures of Heusler alloys have been described in the same $F\bar{4}3m$ space group having both 4c and 4d positions (splitting of $8c$ site in $Fm\bar{3}m$), for a fuller discussion of site comparison.

The calculations of enthalpy of formation, electronic structure and spin polarisation at 0\,K were carried out using density functional theory (DFT). They were conducted using the projector augmented wave (PAW) method~\cite{bl1994hl} implemented in the Vienna \textit{ab initio} simulation package (VASP)~\cite{kresse1994ab,kresse1994norm}.
The exchange-correlation was described by the generalized gradient approximation modified by Perdew, Burke and Ernzerhof (GGA-PBE)~\cite{perdew1996generalized}. Energy bands up to a cutoff of $E$ = 600\,eV were used in all calculations and the convergence tolerance for the calculations was selected as a difference on the total energy within 1 ${\times}$ 10$^{-6}$\,eV/atom. For each structure, volume and ionic (for disordered compounds) relaxation steps were performed and the tetrahedron method with Bl\"ochl correction~\cite{blochl1994improved} was applied.
Spin-polarization calculations were considered for all the structures.
In order to statistically simulate the chemical disorder in Fe$_2$RuGe, unit cells based on the concept of special quasirandom structure (SQS)~\cite{zunger1990special} were generated for different possible disorder schemes.
To generate the SQS, the cluster expansion formalism for  multicomponent and multisublattice systems~\cite{sanchez1984generalized} was used as implemented in the Monte-Carlo (MCSQS) code contained in the Alloy-Theoretic Automated Toolkit (ATAT)~\cite{van2013efficient,van2009multicomponent}.
Subsequent DFT calculations were performed in order to test the quality of the SQS and to see how reliable the DFT results are. The root mean square  (\textit{rms}) error on correlation functions was used as another quality criterion in addition to the calculations including a different order of interactions. The $rms$ error describes the deviation of the correlation function of the SQS ($\Pi^{k}_{SQS}$) from the correlation function of a fully random structure ($\Pi^{k}_{md}$) for all clusters $k$.
\begin{equation}
\ rms = \sqrt{\sum_{k}(\Pi^{k}_{SQS}-\Pi^{k}_{md})^2}
\label{eq1}
\end{equation}

Several tests on the dependence of the type and number of clusters were performed to generate the disordered structure: Ge at 4$a$ (0,0,0), Fe at 4$b$ ($\frac12$,$\frac12$,$\frac12$) and Fe = 0.5/Ru = 0.5 at 4$c$ ($\frac14$,$\frac14$,$\frac14$) and Fe = 0.5 \& Ru = 0.5 at 4$d$ ($\frac34$,$\frac34$,$\frac34$).
Finally, 7 pairs, 5 triplets and 11 quadruplets interactions were considered to obtain reliable results.
Using these parameters, a disordered quaternary SQS cell of 28 atoms was generated.

\section{Results and Discussions}
\subsection {Structural Characterization}
\label{sec:structure}

\begin{figure}[h]
\centering
\includegraphics[width=0.5\textwidth]{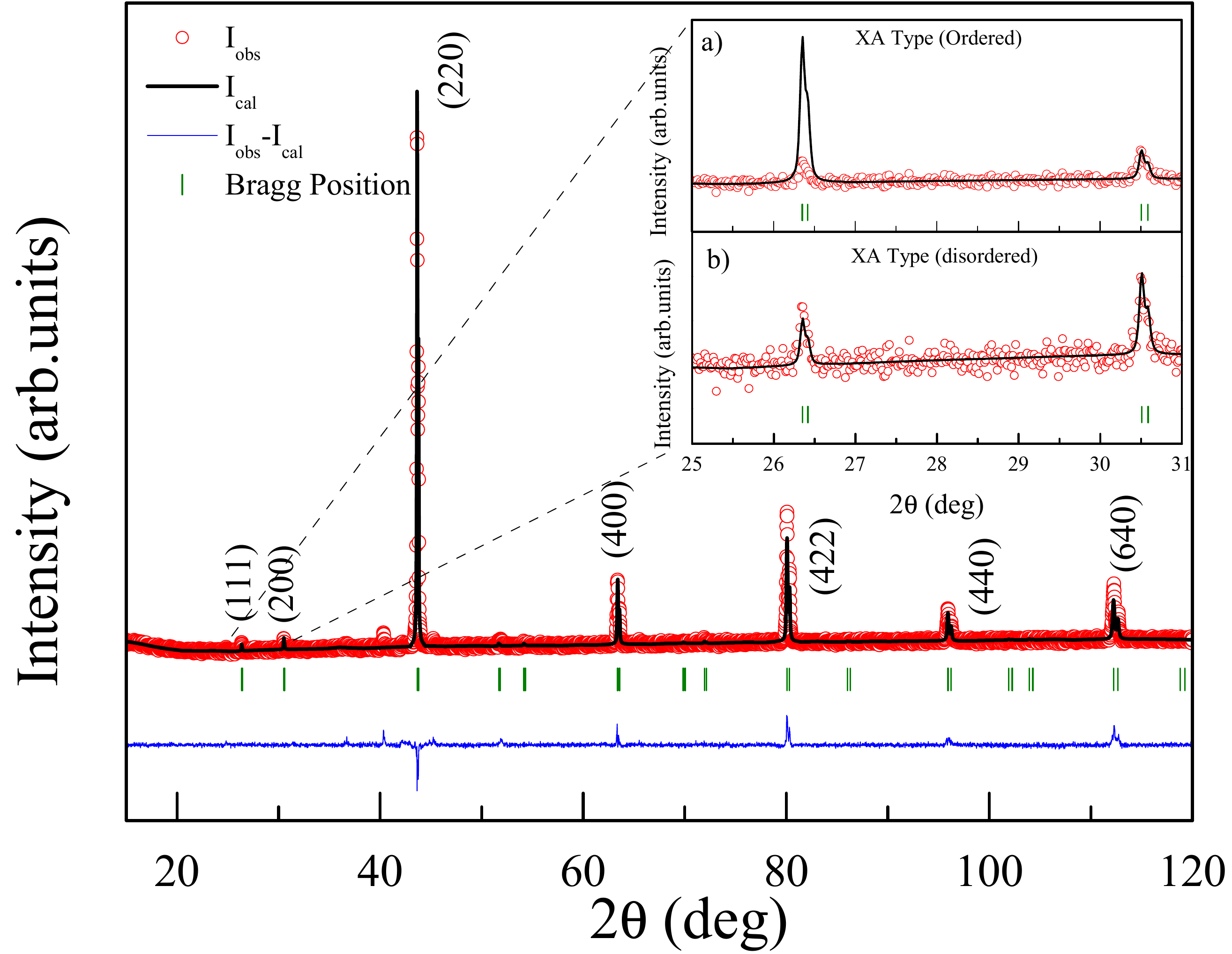}
\caption{Room-temperature powder XRD data along with  Rietveld refinement. Insets (a) and (b) display enlarged views of superlattice reflection peaks fitted by ordered and disordered models, respectively.}
\label{fig:XRD}
\end{figure}

\begin{figure}[h]
\centering
\includegraphics[width=0.5\textwidth]{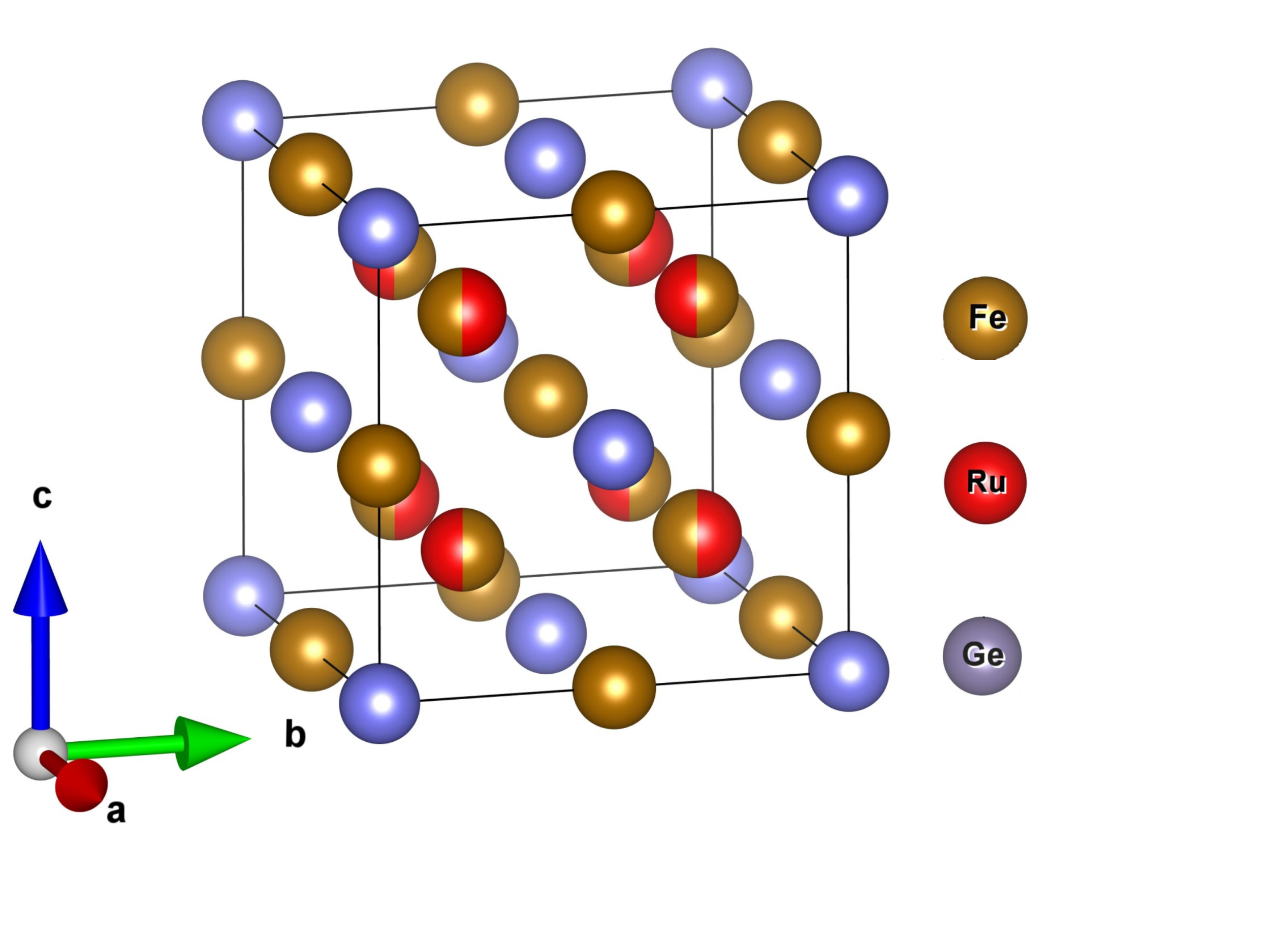}
\caption{Crystal structure of Fe$_2$RuGe in disordered form in $F\bar{4}3m$, mixing $4c$ and $4d$ sites with a 50:50 composition.}
\label{fig:structure}
\end{figure}

The powder XRD data taken at room temperature is presented in Fig.~\ref{fig:XRD}. With the exception of very few small peaks ($<$2\% of main peak intensity), all the peaks in the XRD pattern can be well indexed with a Hg$_2$CuTi-type (space group $F\bar{4}3m$, No - 216) inverse Heusler (XA-type) crystal structure with lattice parameter $a$ = 5.871(4)\,\AA.
It may be noted that Heusler alloys are prone to structural disorder, and despite the formation in cubic structure, such atomic disorders are known to affect the intensity of the (111) and (200) superlattice Bragg peaks in the XRD patterns.
For example, in $A2$-type disorder (all atoms mixed on a single site), both (111) and (200) peaks are absent, while in $B2$-type disorder (atoms mixed on 2 sites), only the (200) peak is present~\cite{mondal2018ferromagnetically}~\cite{gupta2022coexisting}.
The occurrence of both (111) and (200) lines in Fig.~\ref{fig:structure} allows discarding $B2$ or $A2$ type of disorder.
Our attempt to describe the crystal structure with an ordered variant (Ge occupy 4$a$ position, Fe at 4$b$, Fe at 4$c$ and Ru in 4$d$ site) strongly undermined the experimentally observed (111) peak intensity (Fig.~\ref{fig:XRD}, Inset), suggesting a disordered structure of different variant.
Accordingly, we tried various combinations of site-disorders and obtained the best result by considering a 50\% permutation between Fe and Ru in the tetrahedral 4$c$ and 4$d$ sites (Fig.~\ref{fig:structure}).

The Rietveld analysis, considering the above model, can fit the XRD pattern quite well (Fig.~\ref{fig:XRD}). The trace amount of the minority phase turns out to be that of FeGe$_2$ of less than 2\% of the weight fraction of the primary phase. The structural disorder is further confirmed by the EXAFS and neutron diffraction data analysis which will be discussed in later sections.
This result is also in conformity with the theoretical calculations establishing that the disordered XA structure is more stable than the ordered one.

\subsection {EXAFS analysis}
\label{sec:EXAFS}

\begin{figure}[h]
\centering
\includegraphics [width=0.5\textwidth]{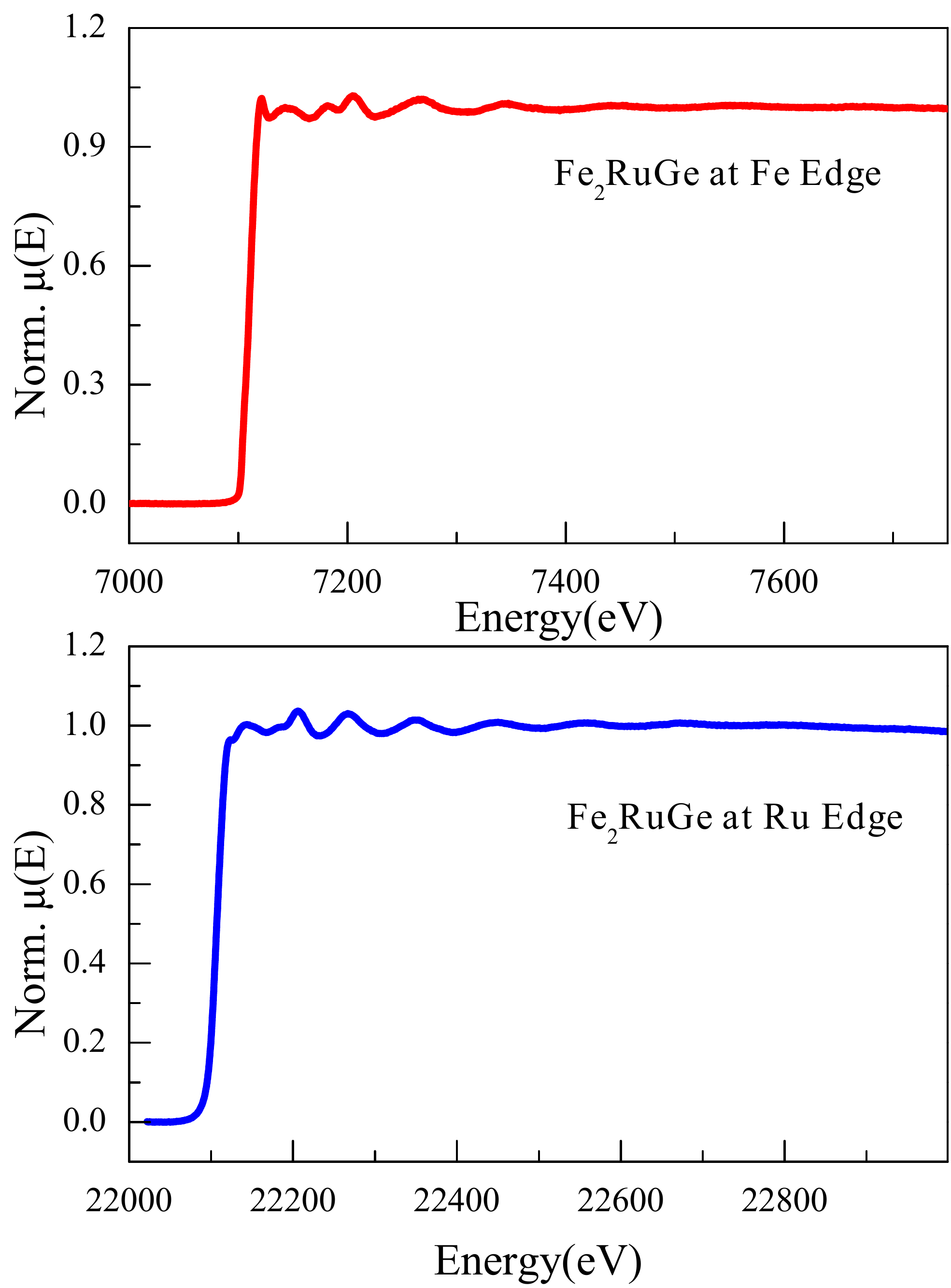}
\caption{Normalized EXAFS spectra of Fe$_2$RuGe taken at (a) Fe edge (b) Ru edge.}
\label{fig:EXAFS}
\end{figure}

\begin{figure}[h]
\centering
\includegraphics [width=0.5\textwidth]{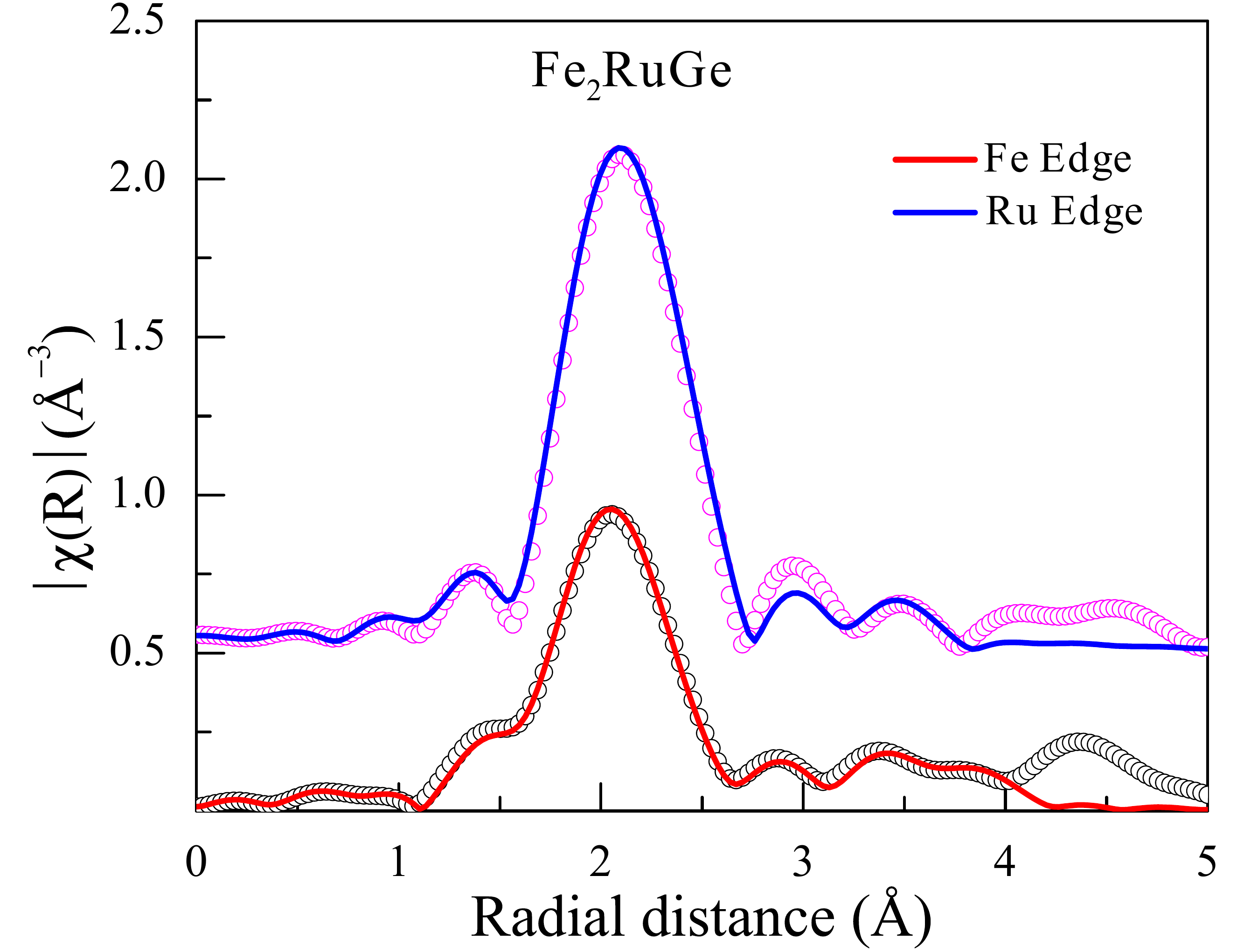}
\caption{Normalized and Fourier transformed EXAFS spectra of Fe$_2$RuGe taken at  Fe edge (upper panel) and Ru edge (lower panel), respectively.}
\label{fig:EXAFS_fourier}
\end{figure}

\begin{table*}[]
\caption{Extracted values of Bond length (R), co-ordination number (N), and Debye-Waller or disorder factor ${\sigma}^2$ obtained
from EXAFS fitting for Fe$_2$RuGe at Fe and Ru edge.}
\begin{tabular}{|cccc|cccc|}
\hline
\multicolumn{4}{|c|}{Fe edge}                                                                                                                                    & \multicolumn{4}{c|}{Ru edge}                                                                                                                                     \\ \hline
\multicolumn{1}{|c|}{Path}   & \multicolumn{1}{c|}{R (\AA)} & \multicolumn{1}{c|}{N}   & ${\sigma}^2$                                             & \multicolumn{1}{c|}{Path}    & \multicolumn{1}{c|}{R (\AA)} & \multicolumn{1}{c|}{N}   & ${\sigma}^2$                                             \\ \hline
\multicolumn{1}{|c|}{Fe-Ru}  & \multicolumn{1}{c|}{2.53 ±0.01}             & \multicolumn{1}{c|}{4}   & \begin{tabular}[c]{@{}c@{}}0.0250\\ ±0.0023\end{tabular} & \multicolumn{1}{c|}{Ru-Ge}  & \multicolumn{1}{c|}{2.53 ±0.01}             & \multicolumn{1}{c|}{4}   & \begin{tabular}[c]{@{}c@{}}0.0082\\ $\pm$0.0003\end{tabular} \\ \hline
\multicolumn{1}{|c|}{Fe-Fe} & \multicolumn{1}{c|}{2.48±0.01}              & \multicolumn{1}{c|}{4}   & \begin{tabular}[c]{@{}c@{}}0.0164\\ ±0.0004\end{tabular} & \multicolumn{1}{c|}{Ru-Fe}  & \multicolumn{1}{c|}{2.53 ±0.01}             & \multicolumn{1}{c|}{4}   & \begin{tabular}[c]{@{}c@{}}0.0082\\ $\pm$0.0003\end{tabular} \\ \hline
\multicolumn{1}{|c|}{Fe-Ge}  & \multicolumn{1}{c|}{2.48±0.01}              & \multicolumn{1}{c|}{3}   & \begin{tabular}[c]{@{}c@{}}0.0164\\ ±0.0004\end{tabular} & \multicolumn{1}{c|}{Ru-Ru} & \multicolumn{1}{c|}{2.88 ±0.01}             & \multicolumn{1}{c|}{2.3} & \begin{tabular}[c]{@{}c@{}}0.0073\\ $\pm$0.0014\end{tabular} \\ \hline
\multicolumn{1}{|c|}{Fe-Ru}  & \multicolumn{1}{c|}{2.89±0.01}              & \multicolumn{1}{c|}{1.7} & \begin{tabular}[c]{@{}c@{}}0.0034\\ ±0.0004\end{tabular} & \multicolumn{1}{c|}{Ru-Fe}  & \multicolumn{1}{c|}{2.88 ±0.01}             & \multicolumn{1}{c|}{3.7} & \begin{tabular}[c]{@{}c@{}}0.0073\\ $\pm$0.0014\end{tabular} \\ \hline
\multicolumn{1}{|c|}{Fe-Fe} & \multicolumn{1}{c|}{2.88±0.01}              & \multicolumn{1}{c|}{4.3} & \begin{tabular}[c]{@{}c@{}}0.0080\\ ±0.0004\end{tabular} & \multicolumn{1}{c|}{Ru-Ru} & \multicolumn{1}{c|}{4.07±0.03}              & \multicolumn{1}{c|}{5.1} & \begin{tabular}[c]{@{}c@{}}0.0075\\ $\pm$0.0047\end{tabular} \\ \hline
\multicolumn{1}{|c|}{Fe-Ge}  & \multicolumn{1}{c|}{3.11±0.01}              & \multicolumn{1}{c|}{6}   & \begin{tabular}[c]{@{}c@{}}0.0247\\ ±0.0017\end{tabular} & \multicolumn{1}{c|}{Ru-Fe}  & \multicolumn{1}{c|}{4.07±0.03}              & \multicolumn{1}{c|}{6.9} & \begin{tabular}[c]{@{}c@{}}0.0075\\ $\pm$0.0047\end{tabular} \\ \hline
\end{tabular}%

\label{tab:EXAFS}
\end{table*}
The normalized EXAFS ($\mu(E)\,vs.\,E$) spectra of Fe$_2$RuGe at the Fe and Ru edges, following the treatment presented in Eq.~\ref{ex1} and Eq.~\ref{eq2}, are shown in Fig.~\ref{fig:EXAFS}.
The $\chi(R)\,vs.\,R$ plots (Fourier transformed EXAFS spectra) generated by the crystallographic information and the EXAFS equation of the Fe and Ru edge is shown in Fig.~\ref{fig:EXAFS_fourier}. Simultaneous fitting with several edges of various data sets is used in this approach. The $\chi(R)\,vs.\,R$ plots calculated at the Ru and Fe edges are fitted at the same time with common fitting parameters.
The statistical significance of the fitting is improved by reducing the number of independent parameters below the Nyquist criterion. The value of  $R_{factor}$ was used to determine the goodness of fit, which is calculated by:
\begin{equation}
R_{factor}=\frac{[Im(\chi_{dat}(r_{i})-\chi_{th}(r_{i})]^{2} + [Re((\chi_{dat}(r_{i})-\chi_{th}(r_{i})]^{2}]^{2}} {[Im(\chi_{dat}(r_i)^{2}]+[Re(\chi_{dat}(r_i)^{2}]}
\label{eq3}
\end{equation}
\noindent
where ${\chi_{th}}$ and ${\chi_{dat}}$ are the theoretical and experimental ${\chi_{R}}$ values, respectively. $Re$ and $Im$ are the real and imaginary components of the respective quantities.
To theoretically simulate the EXAFS spectra of Fe$_2$RuGe, the structural information (lattice parameters and nearest neighbour numbers) have been taken from the XRD results.
The atomic distances (R), coordination numbers (N), and disorder (Debye-Waller) factors (${\sigma^2}$), which give the mean square variations of  the distances, were used as fitting parameters.
The fitting was done up to 4\,{\AA}. A simultaneous fitting of the Ru and Fe edges could be obtained for the disordered structure only. The refined values of the parameters are listed in Table ~\ref{tab:EXAFS}.

From the fitting of Fe edge of the Fourier transformed spectra, we found that the main peak near 2\,\AA\  has a contribution from all Fe-Fe (2.48\,\AA), Fe-Ru (2.53\,\AA) and Fe-Ge (2.48\,\AA) distances. The second intense peak near 3\,\AA\ arises due to the contribution of Fe-Ru (2.89\,\AA) and Fe-Fe (2.89\,\AA) and the peak near 4\,\AA\ consists of the contribution of Fe-Fe (4.07\,\AA) and Fe-Ru (4.17\,\AA).
From the coordination number and bond length analysis, we found occupancy of the $4b$, $4c$ and $4d$ sites between Fe and Ru is nearly 71\%:29\%, close to 66\%:33\%.
Similarly, we analysed the Ru-edge Fourier transformed spectra, where the main peak near 2\,\AA\ arises due to the contribution of scattering paths of Ru-Ge (2.53\,\AA) and Ru-Ru (2.53\,\AA). The second peak consists of the scattering paths of Ru-Ru (2.88\,\AA) and Ru-Fe (2.88\,\AA) and the third peak near 4\,\AA\ consists of Ru-Fe (4.07\,\AA) and Ru-Ru (4.07\,\AA). From these data we also calculated the intermixing between 4\textit{c} and 4\textit{d} sites to be 58:42. This data is consistent with the Fe edge data also. From the XRD analysis in Sec.~\ref{sec:structure} we found that the disorder between Fe and Ru in 4\textit{c} and 4\textit{d} sites is about 50--50. Our EXAFS analysis is also in agreement with this.

\subsection {DC Magnetization Study}
\label{sec:DC magnetization}

To measure the magnetic ordering temperature and magnetic moment of Fe$_2$RuGe, the magnetization as a function of temperature as well as magnetic field was measured.
\begin{figure}[h]
\centering
\includegraphics [width=0.5\textwidth]{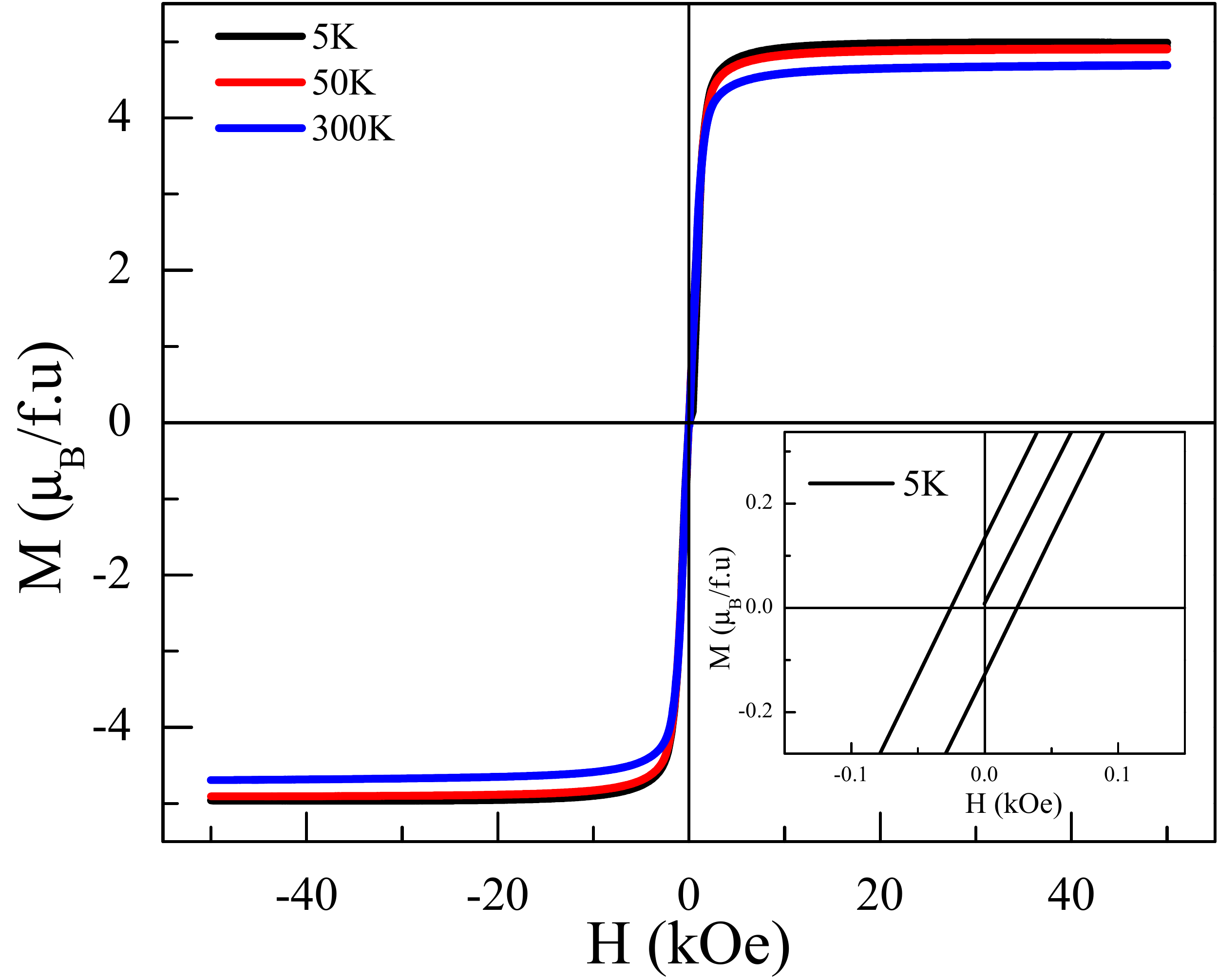}
\caption{Isothermal magnetization plot at various temperatures. Inset represents the hysteresis phenomenon at T = 5\,K}
\label{fig:MH}
\end{figure}
\begin{figure}[h]
\centering
\includegraphics [width=0.5\textwidth]{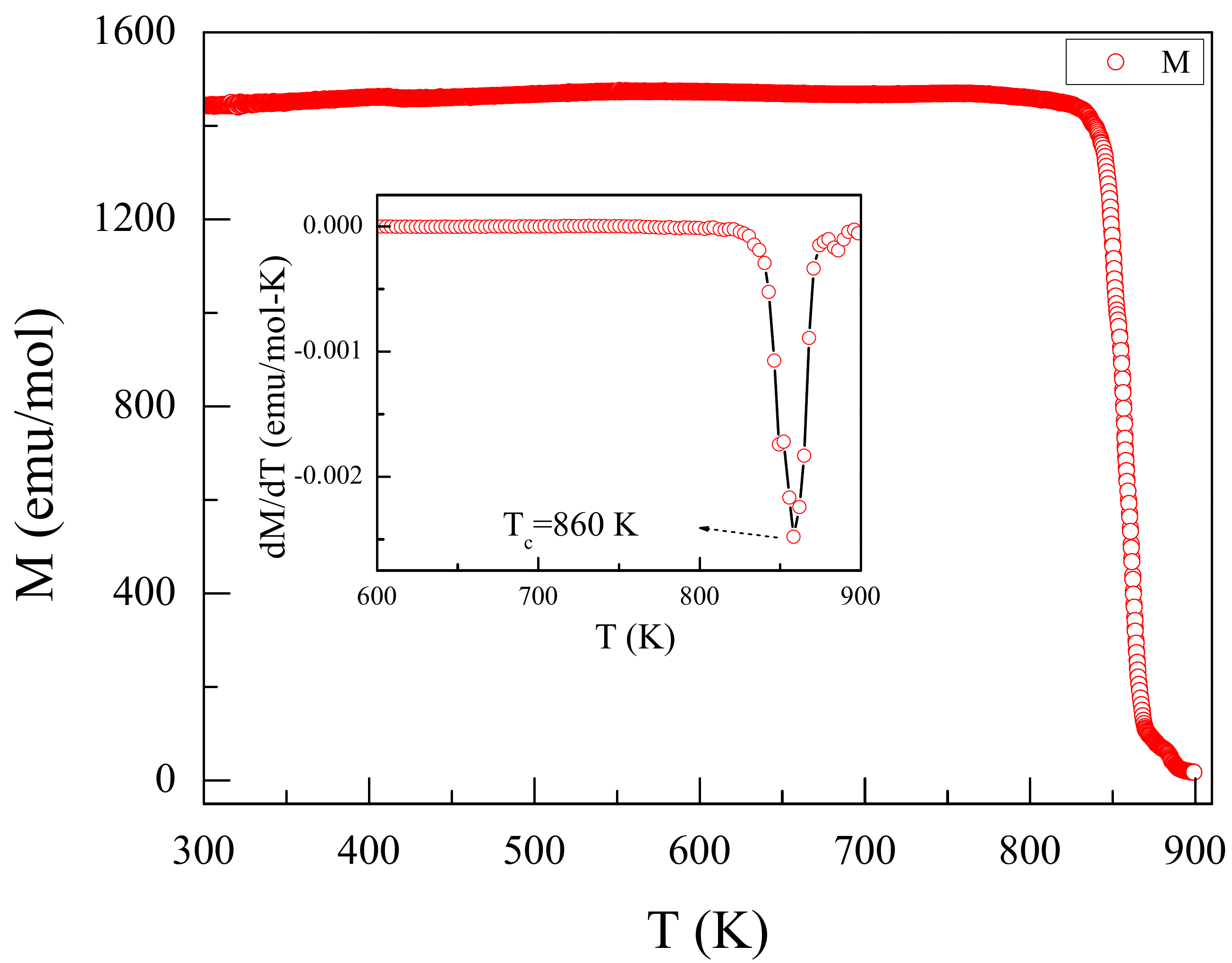}
\caption{ Magnetization \textit{vs}. Temperature plot in 100\,Oe applied magnetic field. Inset shows the first derivative of magnetization as a function of temperature.}
\label{fig:MT}
\end{figure}
The isothermal magnetic measurements at room temperature show a soft ferromagnetic character with a rather large magnetic moment ($\sim$4.71\,$\mu_{\rm B}$/f.u.) (Fig.~\ref{fig:MH}), suggesting ferromagnetic ordering at a temperature above 300\,K.
Accordingly, we performed magnetic susceptibility measurements in a VSM with electromagnet in the temperature range 300 - 900\,K under a magnetic field of 100\,Oe.
Very interestingly, the system exhibits ferromagnetic ordering at $\sim$860\,K (estimated from $dM/dT$ curve) (Fig.~\ref{fig:MT}, Inset), much higher than its binary parent compound, Fe$_3$Ge ($T_{\rm C}$ = 755\,K)~\cite{mcguire2018tuning}.
It may be noted here that the minor phase of FeGe$_2$ detected from our XRD analysis are known to exhibit a ferromagnetic ordering at a much lower temperature, $\sim 290$\,K~\cite{jeong2007electronic}, and hence the observed ordering at $\sim 860$\,K must arise from the main phase. The large magnetic moment in the isothermal magnetic measurement at 300\,K also confirms the ferromagnetic ordering of the primary phase, Fe$_2$RuGe.
Substituting a magnetic Fe atom with a non-magnetic Ru atom, having a random mixed site occupancy  increases \textit{T$_{\rm C}$} even in comparison to the case where 3$d$ atoms like V, Cr, Mn, and Ni are substituted~\cite{mahat2020tuneable,mahat2021influence,keshavarz2019fe2mnge,gasi2013iron}. The isothermal magnetic measurement was also extended to lower temperatures to compare it to the theoretically estimated moments. The data taken at 5, 50 and 300\,K are shown in Fig.~\ref{fig:MH}. Even at the lowest measured temperature, 5\,K, the coercive field retains its soft magnetic character ($H_C$ = 130\,Oe), and the saturation moment is only slightly enhanced  to $\sim$4.92 $\mu_{\rm B}$/f.u.. It may be pointed out here that the well-known Slater-Pauling (SP) rule provides a theoretical limit of the magnetic moments in Heusler alloy systems. According to the SP rule, the saturation magnetic moment ($M$) can be expressed as (N$_V$ -24)$\mu_{\rm B}$/f.u. where N$_V$ is the valence electron number~\cite{graf2011simple}. The atomic disorder present in the system is known to reduce the experimental value in comparison to the SP-rule prediction~\cite{gupta2022coexisting}. Since Fe and Ru are randomly distributed in 4$c$ and 4$d$ positions in Fe$_2$RuGe, one thus expects a considerably diminished value of magnetic moment. Surprisingly, the experimentally observed saturation magnetic moment ($\sim$4.92 $\mu_{\rm B}$/f.u.) for Fe$_2$RuGe is found to be much higher than that theoretically predicted (4\,$\mu_{\rm B}$/f.u.). Although a few Fe-based HA are already known to exhibit saturation moment higher than that predicted by the SP-rule~\cite{gasi2013iron}, Fe$_2$RuGe occupies a special place because it exhibits such a large value despite its inherent atomic disorder involving the moment carrying Fe-atoms.
As up to now all studied HMF systems obey the SP rule~\cite{gupta2022coexisting}, the specific behaviour of Fe$_2$RuGe needs other experimental measurements (neutron diffraction and Mössbauer spectrometry) as well as theoretical calculations to be more clearly  understood.

\subsection {Neutron Diffraction}
\label{sec:ND}

\begin{figure}[h]
\centering
\includegraphics [width=0.5\textwidth]{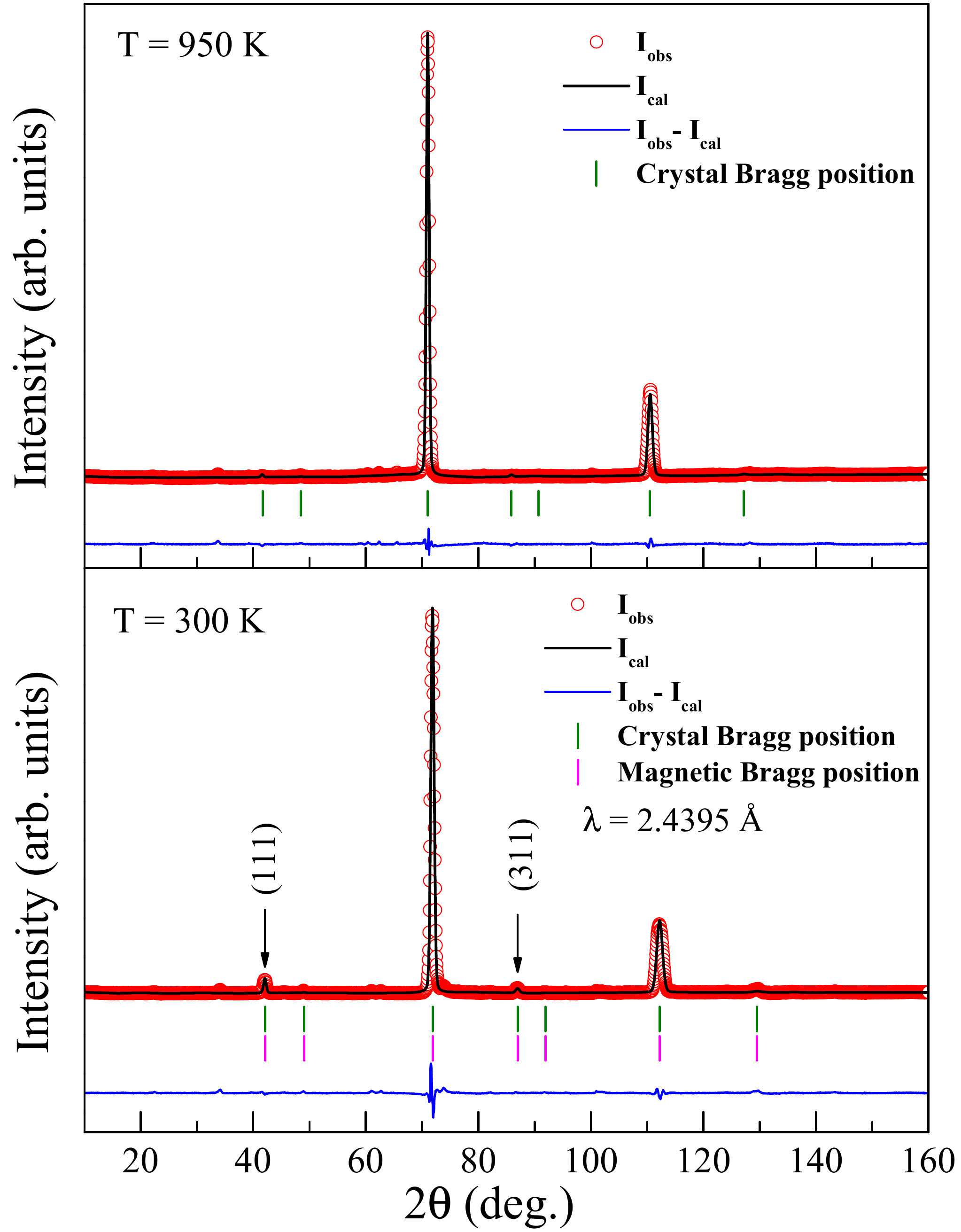}
\caption{Neutron diffraction pattern along with full Rietveld refinement of paramagnetic (upper panel) and magnetically
ordered temperature (lower panel)}
\label{fig:ND}
\end{figure}

Neutron powder diffraction measurements have been carried out on Fe$_2$RuGe at 300\,K and 950\,K, respectively, below and above its Curie temperature ($T_{\rm C}\sim 860$\,K) to analyze the nuclear structure at 950\,K and the long-range magnetic order at 300\,K (Fig.~\ref{fig:ND}). The diffraction pattern at 950\,K, i.e., in the paramagnetic region, can be well refined in the cubic space group $F\bar{4}3m$ with lattice parameter as $a$ = 5.936(2)\,\AA, while the Fe and Ru occupy the 4$c$ and 4$d$ positions with occupation rates of 55\% and 45\%, respectively. This agrees with the XRD data analysis described earlier in Sec.~\ref{sec:structure} and confirms the larger stability of the disordered XA structure compared to Heusler or ordered XA structure. We may, however, note that the intensity of (111) and (200) Bragg peaks, which are generally considered as a signature of perfect crystalline ordering, have much-reduced intensity in the ND spectra in comparison to XRD data, which can be explained due to the different atomic scattering factors of neutron and X-ray. The small peak at 2\,$\theta$ = 33.9° can be refined with 0.6\% of FeGe$_2$ impurity. The NPD pattern measured at 300\,K, in the magnetically ordered state, appears to be nearly like that at 950\,K, except for a moderate increase in the intensity of the (111) and (311) crystal Bragg peaks. Assuming the structural model refined in the paramagnetic region (950\,K) remains similar at 300\,K, the absence of additional peaks beyond the allowed crystal Bragg positions ruled out the antiferromagnetic type of order in this system.
Rietveld refinement of the 300\,K data yields a refined cell parameter $a$ = 5.8787(2)\,\AA\ close to the XRD value. A good refinement was obtained assuming a ferromagnetic alignment of the Fe sublattice in agreement with the magnetization data.
In a powder pattern, all the axes are equivalent to a cubic structure, and it is not possible to specify a preferred orientation of the magnetic moments. Since the intensity of only a few Bragg peaks is related to magnetic ordering, with a small intensity compared to the nuclear contribution, it was not possible to refine independently each Fe moment without a divergence of the fit.
A unique Fe moment was therefore assumed for all sites and the refinement of the Fe moment intensity lead to 2.49(4) $\mu_{\rm B}$/Fe or 4.98 $\mu_{\rm B}$/f.u ($RB_{nucl}$ = 0.65 \% and $R_{Mag}$ =7.78\%). This value agrees quite well with the isothermal magnetization measurements (Sec.~\ref{sec:DC magnetization}) with $M$ = 4.69\, $\mu_{\rm B}$/f.u at 300\,K and $M$ = 4.92\,$\mu_{\rm B}$/f.u at 5\,K.
In order to check if a Ru magnetic contribution had to be taken into account, a fit was performed by fixing the values of Fe and Ru moment to the values calculated by DFT: $M$(Fe-$4b$) = 2.74\,$\mu_{\rm B}$/Fe, $M$(Fe-$4d$) = 1.92\,$\mu_{\rm B}$/Fe and $M$(Ru) = 0.33\,$\mu_{\rm B}$/Ru (See Sec.~\ref{sec:Theory}). This lead to higher values of the refinement’s factors (RB$_{nucl}$ = 1.25\% and R$_{Mag}$ =15\%). To conclude this part, we consider that the larger total Fe moment in the XA ordered or disordered structure is related to the Fe:Ru $4b$ charge transfer.

\subsection {$^{57}$Fe M\"{o}ssbauer Spectrometry}
\label{sec:Mossbauer}

\begin{figure}[h]
\centering
\includegraphics [width=0.5\textwidth]{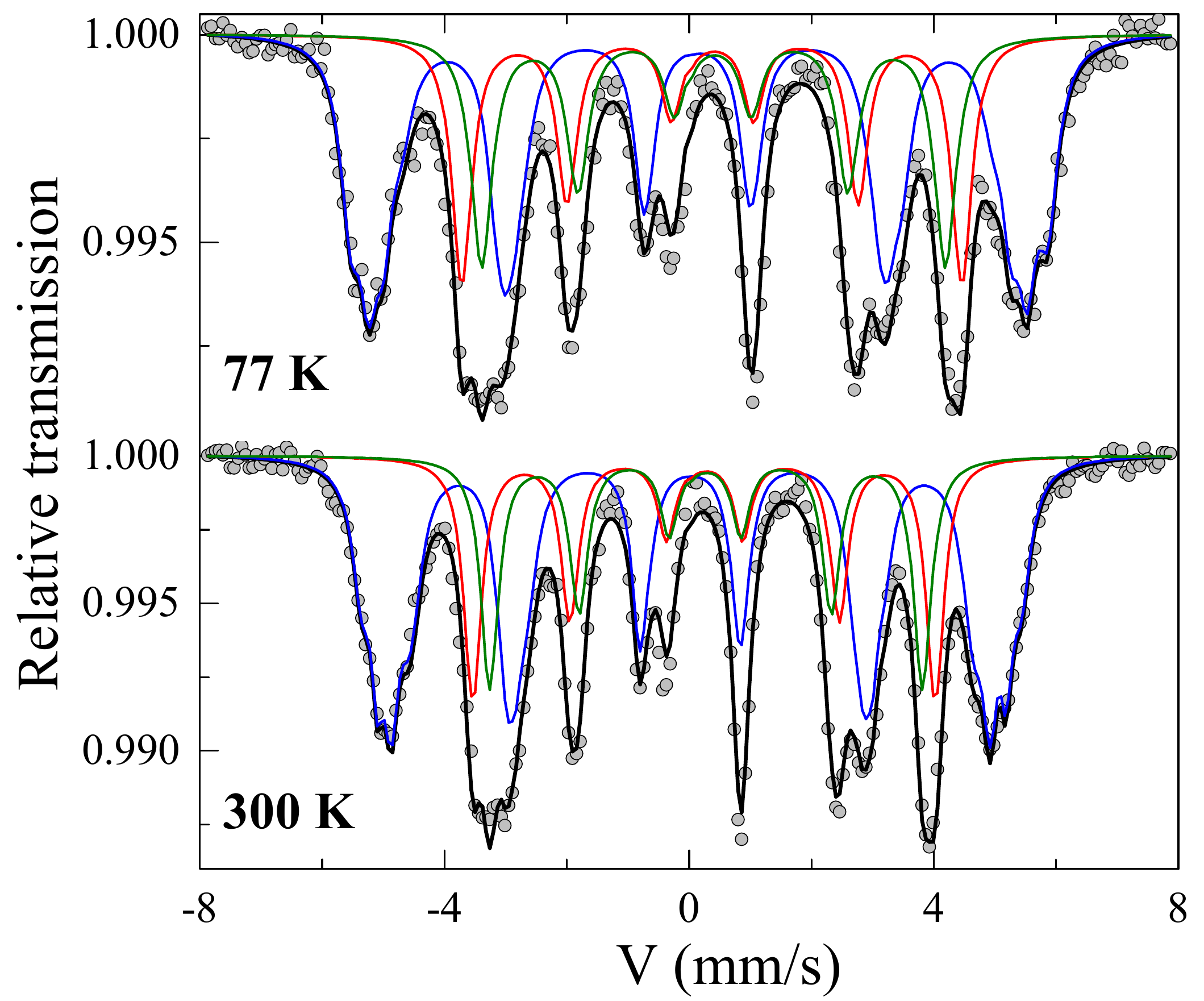}
\caption{$^{57}$Fe M\"{o}ssbauer spectra of Fe$_2$RuGe taken at 300\,K (lower panel) and at 77\,K (upper panel), respectively.}
\label{fig:Mossbauer}
\end{figure}

\begin{table}[]
\caption{Refined values of hyperfine parameters estimated from the M\"{o}ssbauer spectra at 300 and 77\,K: $\delta$, 2$\epsilon$ and B$_{hf}$ correspond to the isomer shift, quadrupolar shift and hyperfine field, respectively.}
\begin{tabular}{|c|c|c|c|c|c|}
\hline
\multicolumn{1}{|l|}{T (K)} & \multicolumn{1}{l|}{Site} & \multicolumn{1}{l|}{\begin{tabular}[c]{@{}l@{}}  $\delta$\\ mm/s\\ $\pm$0.01\end{tabular}} & \multicolumn{1}{l|}{\begin{tabular}[c]{@{}l@{}}2$\epsilon$\\ mm/s\\ $\pm$0.01\end{tabular}} & \multicolumn{1}{l|}{\begin{tabular}[c]{@{}l@{}}B$_{hf}$\\ kOe\\ $\pm$0.5\end{tabular}} & \multicolumn{1}{l|}{\begin{tabular}[c]{@{}l@{}}ratio (\%)\\ $\pm$1\end{tabular}} \\ \hline
\multirow{4}{*}{300\,K}      & Fe-$4b$                       & 0.17                                                                          & 0.01                                                                            & 306                                                                         & 55                                                                      \\ \cline{2-6}
                            & Fe-$4c$                       & 0.39                                                                          & -0.01                                                                            & 233                                                                        & 23                                                                      \\ \cline{2-6}
                            & Fe-$4d$                       & 0.41                                                                          & 0.02                                                                            & 218                                                                         & 22                                                                      \\ \cline{2-6}
                            & \multicolumn{1}{l|}{}     & \multicolumn{1}{l|}{\textless{}0.28\textgreater{}}                            & \multicolumn{1}{l|}{\textless{}0.01\textgreater{}}                              & \multicolumn{1}{l|}{\textless{}269\textgreater{}}                           & \multicolumn{1}{l|}{}                                                   \\ \hline
\multirow{4}{*}{77\,K}       & Fe-$4b$                       & 0.28                                                                          & -0.01                                                                           & 330                                                                         & 52                                                                      \\ \cline{2-6}
                            & Fe-$4c$                       & 0.50                                                                          & 0.01                                                                            & 252                                                                        & 24                                                                      \\ \cline{2-6}
                            & Fe-$4d$                       & 0.53                                                                          & 0.01                                                                            & 234                                                                         & 24                                                                      \\ \cline{2-6}
                            & \multicolumn{1}{l|}{}     & \multicolumn{1}{l|}{\textless{}0.40\textgreater{}}                            & \multicolumn{1}{l|}{\textless{}0.01\textgreater{}}                              & \multicolumn{1}{l|}{\textless{}288\textgreater{}}                           & \multicolumn{1}{l|}{}                                                   \\ \hline
\end{tabular}
\label{tab:Moss}
\end{table}

As the analysis of the Fe$_2$RuGe ND spectra does not allow the determination of site-specific Fe moments, $^{57}$Fe M\"{o}ssbauer spectrometry was used to check the relative population of Fe atoms and their respective magnetic moments (determined appropriately from the magnetic hyperfine fields of the individual components) in the three crystallographic positions, namely 4$b$, 4$c$ and 4$d$. The M\"{o}ssbauer spectra taken at 300\,K and 77\,K (shown in Fig.~\ref{fig:Mossbauer}) are quite similar and show complex magnetic structures consisting of well-resolved magnetic hyperfine structures with broadened and asymmetrical lines. Several models can be proposed to describe each spectrum: the first solution consists of two magnetic components with broad and Lorentzian lines characterized by two different values of isomer shift attributed to the two Fe 4$b$ and 4$d$ sites.
The Fe, Ru and Ge disorder explains why only two components are not able to describe perfectly the hyperfine structures observed at 300 and 77\,K. Indeed, the best fitting model for both spectra at once requires three magnetic components, each resulting from the superposition of several magnetic sextets to describe the asymmetry and width of the lines, in order to take into account the disordered atomic environments. The corresponding refined mean values of the hyperfine parameters are listed in Table~\ref{tab:Moss}.
The relative absorption area is estimated at 55\%, 23\% and 22\% for the three Fe sites.
Assuming the same values of their respective Lamb-M\"{o}ssbauer factors, they directly give rise to the proportions of Fe atoms that match quite well with our analysis of ND and XRD patterns (Sec.~\ref{sec:structure}, Sec.~\ref{sec:ND}). Considering the fact the larger  hyperfine field value corresponds to the 4$b$ site in inverse Heusler alloys~\cite{gasi2013iron}, the mean hyperfine field values at 300\,K for the three magnetic components as 306\,kOe, 233\,kOe and 218\,kOe are assigned to 4$b$, 4$c$ and 4$d$ sites, respectively. The individual Fe moments corresponding to the Wyckoff positions 4$b$, 4$c$ and 4$d$ in Fe$_2$RuGe can be estimated from the three different hyperfine field components, provided that the magnetic hyperfine field values can be properly scaled with the respective magnetic moments.
Usually, $\alpha$-Fe with a magnetic moment of 2.2 $\mu_{\rm B}$ yields a hyperfine field of $\sim$330\,kOe, and most often this reference is used as the calibration factor. Using this approach, it was concluded in a previous study on the structurally similar compound Fe$_{3-x}$Ru$_x$Si that Ru atoms contains a magnetic moment of about $\sim$1.3 $\mu_{\rm B}$ for $x$ = 0.75, while the proposed moment in Ru is slightly reduced to 1\,$\mu_{\rm B}$ in Fe$_2$RuSi~\cite{mishra1985ruthenium}.
This result is really surprising, as only a handful of materials are known where Ru carries magnetic moments~\cite{cao1997thermal}.
If we implement the same calibration factor in Fe$_2$RuGe as well, and compare the result with isothermal magnetisation data, we also would have to assign a magnetic moment of 0.7\,$\mu_{\rm B}$ on Ru atoms in our sample of Fe$_2$RuGe.
However, the neutron diffraction experiment presented above ruled out the presence of any significant magnetic moment at the Ru site. In this context, we may like to point out that the relationship between the magnetic hyperfine field and the magnetic moment are not so straight-forward as elaborated earlier in literature for many different Fe-containing DO$_3$-type alloys~\cite{dubiel2009relationship},~\cite{drijver1976magnetic} as well as for other binary systems~\cite{arzhnikov2001hyperfine}. In some other Fe-based Heusler compounds too, it has also been suggested that a magnetic moment of 2 2\,$\mu_{\rm B}$/Fe would correspond to 255\,kOe~\cite{gasi2013iron}. Such modification is generally caused by the presence of a large anisotropic hyperfine field involving a dipolar field, non-cubic distribution of spin density in the iron atom, an anisotropic g-factor, etc~\cite{drijver1976magnetic}. For our calculation, we have thus utilized the reference of Fe$_3$Ge, which can be regarded as the parent compound of our studied Heusler system, Fe$_2$RuGe, where a magnetic hyperfine field value of 220\,kOe had been associated with 2 $\mu_{\rm B}$/Fe atom~\cite{gasi2013iron}. If we use this calibration factor for Fe$_2$RuGe, we find the Fe magnetic moment values at 4$b$, 4$c$ and 4$d$ sites as 2.78(1), 2.12(1) and 1.98(1)\,$\mu_{\rm B}$, respectively, giving a resulting magnetic moment of $<$4.83(2)$>$ $\mu_{\rm B}$, that matches quite well with the isothermal magnetisation measurement.

\subsection {Theoretical analysis}
\label{sec:Theory}

\begin{figure*}
\centering
\includegraphics[width=1\textwidth]{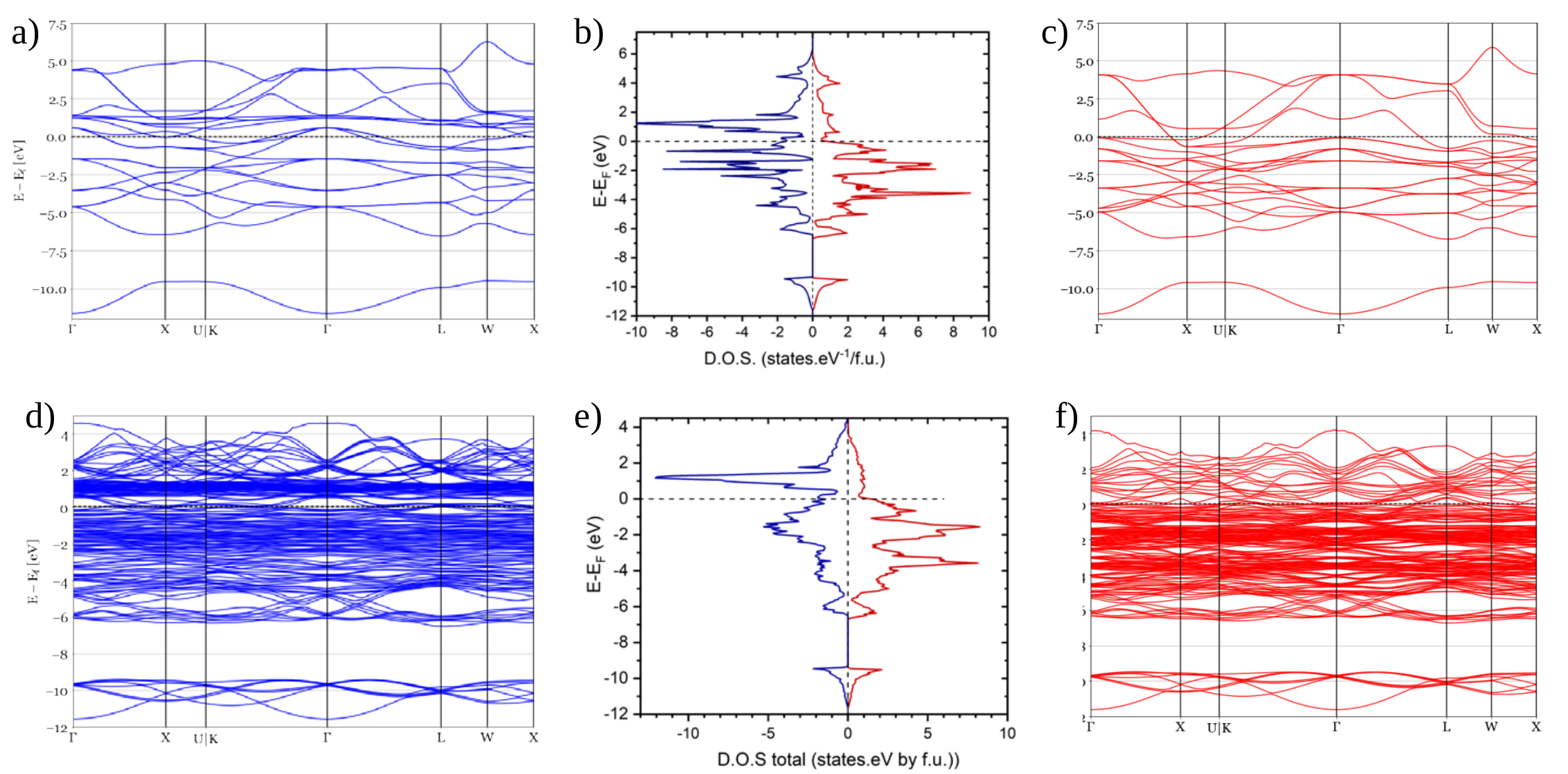}
\caption{Spin-polarized band structure and density of states of Fe$_2$RuGe in ordered Hg$_2$CuTi-type structure: (a) minority (spin-down)
band (b) density of states, (c) majority (spin-up) band. Spin-polarized band structure and density of states of Fe$_2$RuGe in disordered Hg$_2$CuTi-type structure: (d) minority (spin-down)
band (e) density of states, (f) majority (spin-up) band. The energy axis zero point has been set at the Fermi level and the
spin-up (minority) and spin-down (majority) electrons are represented by positive and negative values of the DOS, respectively.}
\label{fig:DFT}
\end{figure*}

In order to theoretically establish the most stable crystal structure, several crystal configurations were studied by DFT calculations.
We considered the ordered Cu$_2$MnAl-type corresponding to the full Heusler structure ($L2_1$) and the Hg$_2$CuTi-type known as the XA structure.
For both structures, the optimized crystal structure was obtained from DFT calculations and the enthalpies of formation are summarized in Table~\ref{Enthalpy}.
Our calculations show that the Cu$_2$MnAl-type structure is less stable than the XA structure which is in good agreement with the structural analysis (Sec.~III.A.).
\begin{table*}[]
\caption{Calculated enthalpy of formation $\Delta_f{H}$ and total magnetic moment for each ordered type of Fe$_2$RuGe, and one disordered case.}
\begin{tabular}{|c|c|c|c|c|c|c|}
\hline & 4\textit{a} & 4\textit{b} & 4\textit{c} & 4\textit{d} & $\Delta_f{H}$ (kJ/mol$^{-1})$ & Magnetic Moment ($\mu_{\rm B}$/f.u.)\\
\hline Heusler (Cu$_2$MnAl)   & Ge & Ru & Fe & Fe  & 2.79  & 5.0  \\
\hline XA structure (Hg$_2$CuTi)   & Ge & Fe  & Ru & Fe & -18.94  & 5.0   \\
\hline Disordered XA structure & Ge & Fe & Fe:Ru  & Fe:Ru & -20.24 & 4.9 \\
\hline
\end{tabular}
\label{Enthalpy}
\end{table*}
Moreover, a structural disorder of Fe and Ru on the 4$c$ and 4$d$ positions, in the XA structure, leads to a more stable compound, thus confirming the XRD, ND and EXAFS data. \\
\begin{table*}[]
\caption{Calculated magnetic moment for each ordered type of Fe$_2$RuGe, and one disordered case.}
\begin{tabular}{|c|c|c|c|c|c|c|c|}
\hline & 4\textit{a} ($\mu_{\rm B}$/f.u.) & 4\textit{b} ($\mu_{\rm B}$/f.u.) & 4\textit{c} ($\mu_{\rm B}$/f.u.) & 4\textit{d} ($\mu_{\rm B}$/f.u.) & $P$ (\%) \\
\hline $L2_1$ (Cu$_2$MnAl)   & Ge = -0.05 & Ru = 1.3  & Fe = 1.94  & Fe = 1.94 & -33.3\,\% \\
\hline XA  (Hg$_2$CuTi)   & Ge  = -0.04  & Fe = 2.76 & Ru  = 0.33  & Fe  = 1.92 & -29.9\,\%  \\
\hline Disordered XA  & Ge  = -0.04& Fe  = 2.74 & Fe:Ru with  Ru  = 0.16& Fe:Ru with Fe =1.98 & -2.7\,\%\\
\hline
\end{tabular}
\label{Magnetic}
\end{table*}
Fig.~\ref{fig:DFT}(a-c) shows the calculated spin-polarized band structure and the calculated density of states (DOS) of the ordered XA structure. As it can be seen, the DOS shows no band gap at the Fermi level (\textit{$E_{\rm F}$}) for either the majority or minority spin population. This explains a rather low value of the spin polarization, with $P(\rm{XA})=\frac{\rm{DOS}^\uparrow (E_{\rm F})- \rm{DOS}^\downarrow (E_{\rm F})}{\rm{DOS}^\uparrow (E_{\rm F})+ \rm{DOS}^\downarrow (E_{\rm F})}$=$-$29.9\%, and $P$(Heusler structure)=$-$33.3\%.
As in many other Heusler compounds, the spin polarization drastically decreases when the disorder increases, $P$ drops to -2.7\% in the disordered XA inverse Fe$_2$RuGe too.
As in Fig.~\ref{fig:DFT}(a-c), Fig.~\ref{fig:DFT}(d-f) shows the calculated spin-polarized band structure and the associated DOS of the disordered XA structures where no semiconducting behaviour can be noticed because both spin directions correspond to a metallic ground state.
The calculated total magnetic moment, for all configurations, is in good agreement with the experimental magnetic moment, always around 5\,$\mu_{\rm B}$/f.u..
Although all calculated total magnetic moments are almost the same for the three configurations, only the ordered and disordered XA types show significant differences in atomic moment of Fe compared to the full Heusler configuration.
It may be noted that the magnetic moment is shared between the Fe and Ru elements for the Heusler type.
In contrast to this, in the XA structure type, the magnetic moment of the Ru element is considerably decreased whereas the magnetic moment carried by the Fe atom in 4$b$ positions is strongly increased (Table~\ref{Magnetic}).
\begin{figure}
\centering
\includegraphics[width=0.5\textwidth]{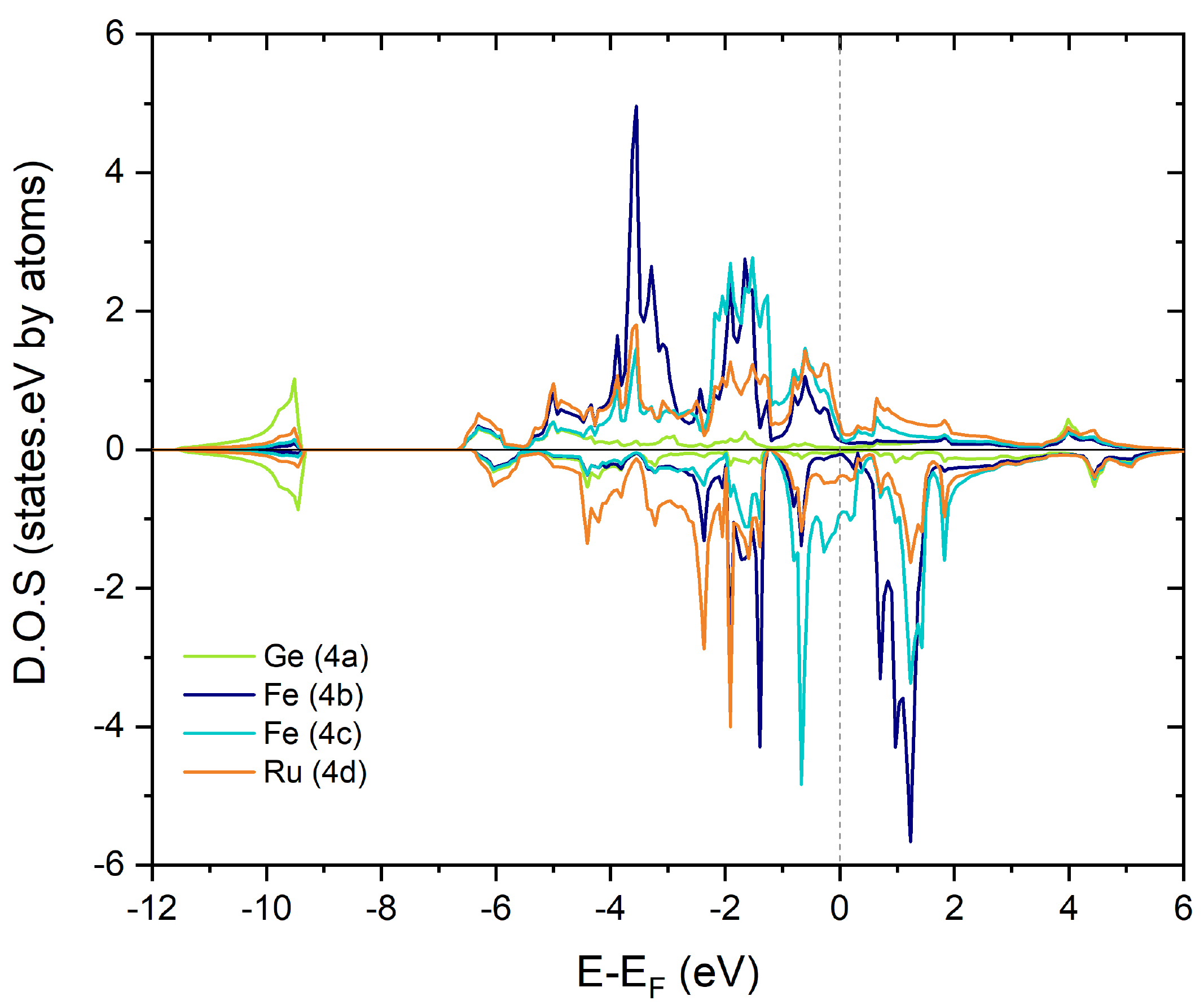}
\caption{Projected SP DOS on each site for the XA-ordered structure.}
\label{fig:PartialDOS}
\end{figure}
In the full Heusler configuration, the Fe-$8c$ site is surrounded by 4~Ge and 4~Ru atoms, while the nearest Fe neighbours are at a large distance and the mean Fe moment is 1.98\,$\mu_\mathrm{B}$/Fe.
In the ordered and disordered XA structures, it can be observed that Fe-$4b$ atoms surrounded by 4~Fe and 4~Ru atoms have a enhanced moment close to 2.8\,$\mu_\mathrm{B}$/Fe while Fe-$4c$($d$) atoms surrounded by 4~Fe and 4~Ge atoms retain a moment close to 1.9\,$\mu_\mathrm{B}$/Fe.
The spin-polarized DOS projected on each site for the XA ordered structure clearly shows a different contribution of the Fe site to the SP-DOS (Fig.\ref{fig:PartialDOS}), with a stronger interaction of Ru orbitals with Fe-$4b$ than with Fe-$4c$.
The calculation of the Bader charges, as shown in Table~\ref{Bader}, reveals a larger electronic charge transfer from Fe-$4b$ (-0.35) to Ru-$4d$ (+0.47) in the disordered XA structure compared to the direct structure, -0.13 to +0.26, for both Fe and Ru sites, respectively.
In fact, the charge transfer from Fe-$4d$ is less important in XA structure since its atomic environment is surrounded by Fe and Ge.
Due to the charge transfer from Fe to Ru, the number of unpaired electrons for the Fe-$4b$ atom increases, explaining the enhancement of its magnetic moment in comparison with Fe-$4d$ and the low value for the Ru moment.
Such charge transfer is also responsible for the enhancement of the magnetic moment of the Fe at $4b$ site, compared to the average moment on Fe in the parent compound Fe$_3$Ge.
To our knowledge, such a large enhancement of magnetic moment has not been reported earlier in any Heusler alloy system.

\begin{table*}[]
\caption{Bader charge transfer in Fe$_2$RuGe}
\begin{tabular}{|c|c|c|c|c|}
\hline & 4\textit{a} (electron) & 4\textit{b} (electron) & 4\textit{c} (electron) & 4\textit{d} (electron) \\
\hline Heusler (Cu$_2$MnAl)   & Ge: +0.00 & Ru: +0.26 & Fe: -0.13 & Fe: -0.13  \\
\hline XA structure (Hg$_2$CuTi)   & Ge: -0.11 & Fe: -0.35  & Ru: +0.53 & Fe: -0.06   \\
\hline Disordered XA structure & Ge: -0.07 & Fe: -0.35 & Ru: +0.47 & Fe: -0.06\\
\hline
\end{tabular}
\label{Bader}
\end{table*}

\subsection {Resistivity study}
\label{sec:resistivity}
\begin{figure}[h]
\centering
\includegraphics [width=0.5\textwidth]{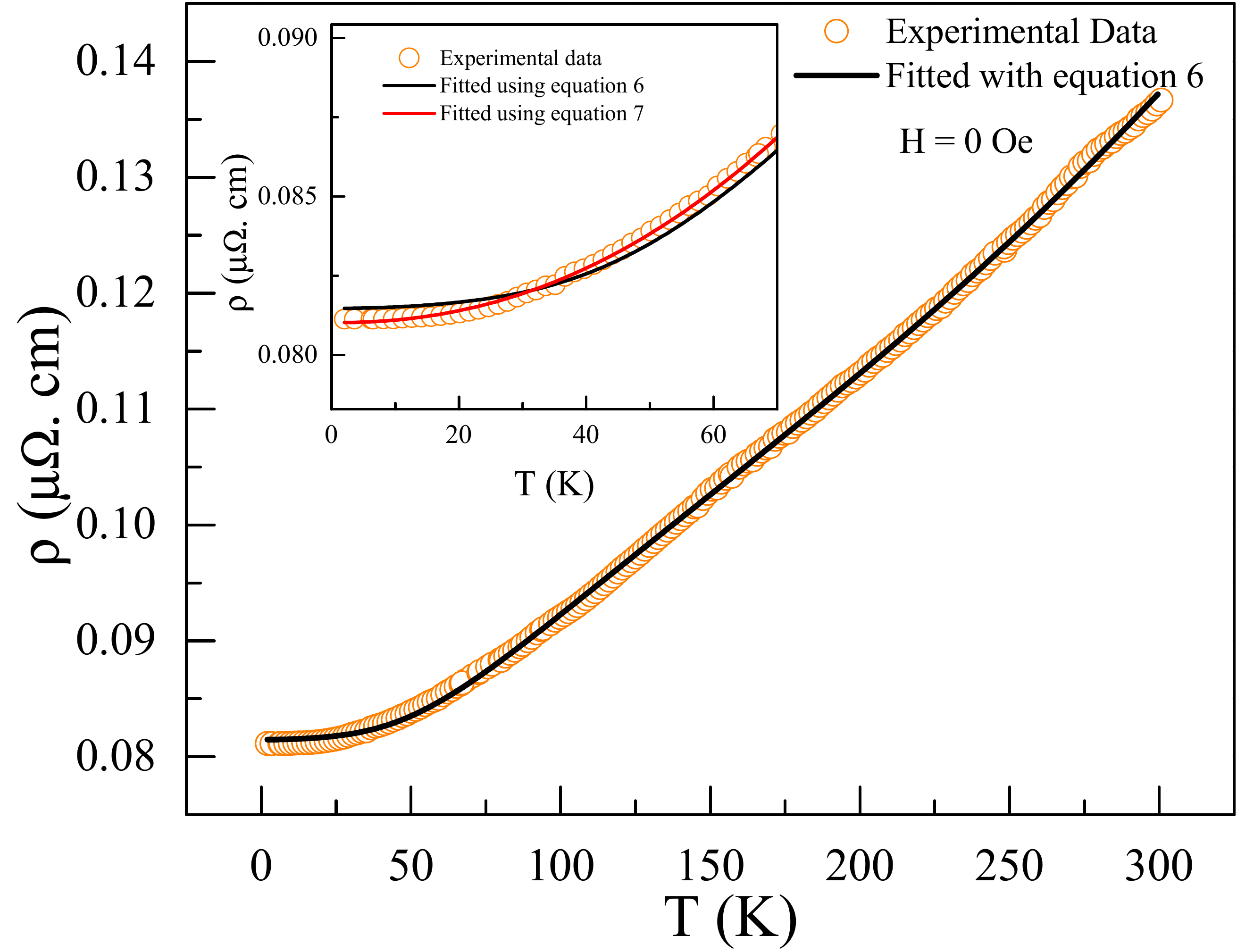}
\caption{Zero field electrical resistivity measured at a function of temperature.}
\label{fig:RT}
\end{figure}

In sec.~\ref{sec:Theory}, it is predicted that Fe$_2$RuGe has density of states in both up and down spin configuration at the Fermi level.
To check its validity, we have performed resistivity experiments.  Fig.~\ref{fig:RT} represents the temperature-dependent electrical resistivity ($\rho(T)$) behaviour of Fe$_2$RuGe measured in the absence of magnetic field ($H$ = 0\,kOe) in both heating and cooling cycles.
We could not observe any irreversibility between these data: thus we ruled out any possibility of a structural transition in the system.
In the case of a ferromagnetic material, the temperature dependence of electrical resistivity can be attributed to three different types of scattering factors.
According to Matthiessen rule, all these contributions arise due to different types of factors such as, a) contribution due to scattering of conduction electrons by the lattice defects, $\rho_0$, b) phonon scattering,  ${\rho}_{P}$, c) magnon scattering and electron--electron interaction, ${\rho}_{M}$, are additive in nature.
From Fig.~\ref{fig:RT} we can see that the residual resistivity ratio (RRR) of Fe$_2$RuGe (${\rho}_{300 K}$/${\rho}_{5 K}$) is around 1.625, which clearly indicates the existence of substantial structural disorder in the sample~\cite{boeuf2006low,nag2022cofevsb}. We can say that resistivity of any ferromagnetic material is
 \begin{equation}
\rho(T) = \rho_0 + \rho_{P}(T) +\rho_{M}(T)
\label{eq:RT}
\end{equation}
\noindent
 The phonon scattering term can be written as~\cite{gupta2022coexisting}~\cite{rossiter1991electrical}
 \begin{equation}
\rho_P = C{\bigg(\frac{T}{\Theta_{D}}\bigg)}^5 \int_{0}^{\frac{\Theta_{D}}{T}} \frac{x^{5}}{(e^{x}-1)(1-e^{-x})} dx
\label{BG}
\end{equation}
\noindent
 and the magnon and electron contribution as $BT^{2}$. Here $\Theta_{\rm{D}}$ is the Debye temperature and $C$ is scattering constant~\cite{gupta2022coexisting}. As the $T_{\rm C}$ is quite high, we have fitted the data using Eq.~\ref{eq:RT}
From the fitted parameters, phonon contribution was found much higher compared to the magnon contribution. However, Eq.~\ref{eq:RT} fits the data well within the region 60 $<$ T $<$ 300\,K only, but is unable to fit the low temperature region properly. From the fitted results we get $\Theta_{\rm{D}}$ = 360\,K. Generally, electron-electron (e-e) and electron-magnon (e-m) lead to a quadratic temperature dependence of resistivity. However, between these two, the dominant contribution can be inferred from the magnitude of the coefficient, $B$. Normally, $B$ is of the order of 10$^{-2}$ n$\Omega$cm.K$^2$ for e-e interaction while it is normally two orders higher for e-m interaction. From the fitting, we got the value of $B$ (e-e interaction coefficient) to be 4.57${\times}10^{-7}$ $\mu\Omega$cm.K$^2$ which falls within the accepted range for electron-electron scattering, indicating the very weak presence of a single-magnon-scattering mechanism.
We have tried to fit the low temperature resistivity data by using the equation

\begin{equation}
\rho(T) = \rho_0 + {\rho}T^n
\label{eq2}
\end{equation}
\noindent
 From the fitting, we have found the value of $n = 2.2$, from which we can confirm that at low temperatures also there is a dominant magnon contribution, which is quite surprising and calls for further investigations of magnetic dynamics in this material.
 Previously, several authors have obtained different power ($n$) values depending on the temperature range considered. The value of $n$ above 2 gives some signs of the weak-metallic character even at higher temperatures~\cite{gupta2022coexisting,nag2022cofevsb}.

\section {Concluding Remarks}
\label{sec:conclusion}

To summarize, we have synthesized a new Fe-based Heusler alloy Fe$_2$RuGe through the arc melting technique.
The combined XRD, ND and EXAFS study reveals that the material crystallizes in a disordered structure, which is also supported by the lower formation ground state energy of the disordered structure over the ordered one, calculated by SQS+DFT.
$^{57}$Fe M\"{o}ssbauer spectrometry confirms the inverse Heusler XA disorder with different sextets, strongly suggesting different Fe positions.
From neutron diffraction, we confirm that the disorder between Fe and Ru in the $4c$ and $4d$ sites is 50:50 in nature.
This leads to the mixing of Fe and Ru in the same plane.
All these results validate that there are two different Fe environments compared to the Heusler structure.
The compound exhibits a ferromagnetic to paramagnetic transition at a significantly high Curie temperature of 860\,K and the saturation moment at 5\,K is estimated to be 4.90\,$\mu_B$/f.u., which is higher than the expected SP value of 4\,$\mu_B$/f.u..
Such enhancement of the magnetic moment above the predicted SP value is observed for the first time.
The neutron diffraction experiments reveal that only Fe, distributed over different sites, contributes to the net magnetic moment while no moment could be detected on Ru site.
Both theoretical calculation and the Mössbauer experiment support that  there is a larger moment on Fe-$4b$ site, in comparison with the other disordered Fe-$(4c-4d)$ sites.
This enhanced magnetic moment is explained by a large electronic charge transfer from Fe-$4b$ to Ru, as calculated by DFT.
The resistivity experiment confirms the presence of electron--electron behaviour at the high-temperature range and a weak phonon contribution at low temperature compared to the magnon contribution.

\begin{acknowledgments}
S.C. and S.G. would like to sincerely acknowledge UGC, India,
 and SINP, respectively, for their fellowship.
DFT calculations were performed using HPC resources from
GENCI-CINES (Grant No. 2021-A0100906175). Work at
the Ames National Laboratory (in part) was supported by
the Department of Energy–Basic Energy Sciences, Materials
Sciences and Engineering Division, under Contract No. DEAC02-
07CH11358.
\end{acknowledgments}
\normalem
\bibliographystyle{apsrev4-2}
\end{document}